\PassOptionsToPackage{table}{xcolor}

\documentclass[acmsmall]{acmart}

\renewcommand\footnotetextcopyrightpermission[1]{}

\usepackage[utf8]{inputenc}
\usepackage{algorithmic}
\usepackage{color,soul}
\usepackage{graphicx}
\usepackage{textcomp}
\usepackage{xcolor}
\usepackage{wrapfig}
\usepackage{url}
\def\BibTeX{{\rm B\kern-.05em{\sc i\kern-.025em b}\kern-.08em
    T\kern-.1667em\lower.7ex\hbox{E}\kern-.125emX}}

\usepackage{colortbl} 

\definecolor{lightgray}{RGB}{192,192,192} 
\definecolor{lightergray}{RGB}{239,239,239} 

\begin{document}

\title{Who Speaks for Ethics? How Demographics Shape Ethical Advocacy in Software Development}

\author{Lauren Olson}
\affiliation{%
  \institution{Vrije Universiteit Amsterdam}
  \department{Software and Sustainability}
  \city{Amsterdam}
  \country{The Netherlands}
}
\email{l.a.olson@vu.nl}

\author{Ricarda Anna-Lena Fischer}
\affiliation{%
  \institution{Vrije Universiteit Amsterdam}
  \department{Software and Sustainability}
  \city{Amsterdam}
  \country{The Netherlands}
}
\email{r.a.l.fischer@vu.nl}

\author{Florian Kunneman}
\affiliation{%
  \institution{Utrecht University}
  \city{Utrecht}
  \country{The Netherlands}
}
\email{f.kunneman@uu.nl}

\author{Emitzá Guzmán}
\affiliation{%
  \institution{Vrije Universiteit Amsterdam}
  \department{Software and Sustainability}
  \city{Amsterdam}
  \country{The Netherlands}
}
\email{e.guzmanortega@vu.nl}

\begin{abstract}
The integration of ethics into software development faces significant challenges due to market fundamentalism in organizational practices, where profit often takes precedence over ethical considerations. Additionally, the critical influence of practitioners' individual backgrounds on ethical decision-making remains underexplored, highlighting a gap in comprehensive research. This is especially essential to understand due to the demographic imbalance in software roles. This study investigates ethical concerns in software development, focusing on how they are perceived, prioritized, and addressed by demographically different practitioners. By surveying 217 software practitioners across diverse roles, industries, and countries, we identify critical barriers to ethical integration and examine practitioners’ capacity to mitigate these issues. Our findings reveal pronounced demographic disparities, with marginalized groups—including women, BIPOC, and disabled individuals—reporting ethical concerns at higher frequencies. Notably, marginalized practitioners demonstrated heightened sensitivity to ethical implementation and greater empowerment to address them. However, practitioners overall often lack the support needed to address ethical challenges effectively. These insights underscore the urgent need for reforms in software education and development processes that center on diverse perspectives. Such reforms are essential to advancing ethical integration in software development and ensuring responsible computing practices in an increasingly complex technological landscape.
\end{abstract}

\maketitle

\section{Introduction}
The integration of ethical considerations into software development has garnered significant attention in recent years, reflecting growing concerns about the societal and individual impacts of technology. As ethical frameworks and codes of conduct such as the ACM Code of Ethics attempt to guide practitioners, the practical application of these principles remains challenging~\cite{mcnamara2018does}. Studies reveal that such frameworks often lack actionable guidance~\cite{mcnamara2018does, gogoll2021ethics}, limiting their influence on day-to-day decision-making and leaving organizations vulnerable to "ethics-washing"—using ethical language to deflect criticism without substantive change~\cite{metcalf2019owning, green2021contestation}. Moreover, the homogeneity of the tech workforce exacerbates these challenges, as underrepresentation of women, BIPOC (Black, Indigenous, and People of Color), and lower-income individuals limits the diversity of ethical perspectives~\cite{jakesch2022different, EEOC2024}. This is concerning as practitioners frequently rely on individual initiative to address ethical concerns, a dynamic that is unsustainable without institutional support~\cite{widder2023s, bamberger2015privacy}. 

These limitations highlight the urgent need for comprehensive approaches to ethics integration that account for organizational culture, demographic diversity, and systemic power dynamics. This study provides a starting point to evaluate which aspects of ethics integration require additional demographic consideration either in policy, practice or research. By centering on the perspectives of practitioners and marginalized groups, our research aims to inform the design of future interventions that empower ethical decision-making in software development and advance the field of responsible computing.

This work investigates ethical concerns and their integration into software development through the lens of practitioners’ experiences, offering insights into a diverse population spanning multiple countries, industries, and roles. Unlike much previous research that narrowly focuses on ethical issues related to artificial intelligence (AI)~\cite{jakesch2022different, pant2024ethics} or perspectives of ethics advocates~\cite{widder2023s, wong2021_timelines} and UX designers~\cite{gray2019ethical, wong2021_uxmemos, wong2021_tactics}, our study examines both abstract principles (e.g., transparency, accountability) and concrete challenges (e.g., misinformation, censorship) across a broader spectrum of software practitioners and systems. Drawing on a survey of 217 software practitioners, this study addresses three key research questions:

\begin{minipage}{\linewidth}
\vspace{0.5em}
\textbf{(RQ1)} What ethical concerns do software practitioners perceive to be associated with their software, and what ethical concerns do they think users have regarding their software?

\vspace{0.5em}
\textbf{(RQ2)} Which factors do practitioners think influence the integration and prioritization of ethics during software development?

\vspace{0.5em}
\textbf{(RQ3)} To what degree do software practitioners feel they are able to mitigate ethical concerns in the products they create?
\vspace{0.5em}
\end{minipage}


Our study makes several contributions to the literature. First, it confirms and extends previous findings on the critical role of organizational culture in shaping ethical practices, providing evidence from a diverse population of practitioners. Second, it broadens the scope of ethical concerns by ethical issues from user feedback. Third, it examines how demographic factors influence practitioners’ ability to prioritize and integrate ethics into software, offering an exploratory analysis of marginalized groups, including women, BIPOC, and Global South practitioners. Finally, it underscores the importance of examining ethics within diverse organizational settings, including tech companies and non-tech firms engaged in software development, moving beyond the narrow focus on AI ethics. To encourage further research we make a replication package available~\footnote{\url{https://doi.org/10.6084/m9.figshare.27195492}}.

\section{Background}

When practitioners engage with ethics during development,  multiple levels of influence including organizational norms and individual backgrounds on ethical principles are involved~\cite{gray2019ethical}. These intertwined factors form the foundation for understanding how ethics is integrated into software design and development.

\subsection{Organizational Norms}
Market fundamentalism in tech, where ``market success trumps ethics," poses challenges to integrating ethics~\cite{metcalf2019owning}. In AI ethics, Madaio et al.~\cite{madaio2020co} found that fairness efforts rely on ad-hoc processes driven by individual advocates but are undermined by organizational culture. Effective responsible AI integration requires alignment with workflows, supportive structures, leadership, and advocacy~\cite{rakova2021responsible}. Pant et al.~\cite{pant2024ethics} highlight workplace policies' role in shaping AI ethics awareness, while Ali et al.~\cite{ali2023walking} identify product-launch priorities as barriers for AI ethics workers. Career concerns and unsupportive cultures further weaken ethics advocacy~\cite{widder2023s, rakova2021responsible, madaio2020co}.

Nafus~\cite{nafus2006real} notes that corporate user researchers often align with engineering and marketing paradigms, underscoring tensions between ethics and organizational norms. This paradigm is wide-reaching; ethics codes like the ACM Code of Ethics are found to have limited influence on decision-making~\cite{mcnamara2018does} and codes of conduct provide insufficient practical guidance~\cite{gogoll2021ethics}. Further, ethical frameworks often focus on formalized processes that create an illusion of accountability. Metcalf et al.~\cite{metcalf2019owning} argue these processes reinforce ``imagined objectivity"~\cite{benjamin2019race} while obscuring genuine accountability. Checklist-style procedures and compliance structures can lead to ethics-washing, where ethics language diffuses criticism without meaningful change~\cite{green2021contestation}. Tech ethics therefore often lacks rigor and serves corporate incentives to maintain legitimacy~\cite{green2021contestation}. These findings reveal systemic barriers to ethics advocacy in technology organizations and highlight the need for research into how culture and structure affect ethical decision-making.

These previous studies, elucidating the relationship between organizational norms and ethics integration, often lack transparency regarding demographic data~\cite{madaio2020co, mcnamara2018does}, exhibit a focus on AI-specific practices and professionals~\cite{ ali2023walking, rakova2021responsible, pant2024ethics, kapania2022because}, or center predominantly on the perspectives of ethics advocates~\cite{widder2023s, wong2021_tactics}. Our study builds upon their contributions by expanding the scope beyond AI and ethics advocacy, while also incorporating an analysis of demographic factors to underscore the critical role of individual background when forming ethical perspectives.

While our research builds on Widder et al.~\cite{widder2023s} and Madaio et al.~\cite{madaio2020co} by identifying disparities in ethics handling within software development, our study differs in both scope and goals. Madaio et al. focused on improving AI fairness checklists’ efficacy. Our study instead addresses a broad range of ethical concerns across software applications, allowing us to examine how workplace structures shape the treatment of ethical concerns more generally, rather than concentrating on AI-specific mechanisms.

The focus of Widder et al.’s work and ours also fundamentally diverges. Our study categorizes ethical concerns by type—such as privacy, accessibility, accountability, transparency, and sustainability—using Tjikhoeri et al.’s ethical issues framework~\cite{tjikhoeri2024best}, which supports cross-context comparison and reveals internal organizational challenges, such as companies encouraging excessive data collection, struggling to keep up with legal changes, or deprioritizing accessibility due to budget constraints. By contrast, Widder et al. survey ethics advocates, leave ethical concerns intentionally undefined, and classify concerns by industry or system type, focusing on broader societal harms like military applications, surveillance, and manipulative advertising. Our study, in contrast, offers an organization-centered perspective, using a structured framework to generate empirical insights into ethics integration across a broad population of practitioners.

\subsection{Individual Background}
Research shows that practitioners’ backgrounds significantly shape how they approach ethics in technology development. Widder et al.~\cite{widder2023s} emphasize that addressing ethical concerns often relies on practitioners' personal initiative to overcome organizational barriers, while Bamberger et al.~\cite{bamberger2015privacy} highlight the critical role of ethics officers' individual commitment over regulatory compliance.  

Practitioners’ demographic backgrounds and experiences influence their ethical priorities. Values are shaped by lived experiences~\cite{fumagalli2010gender, grgic2018human}, but the tech workforce remains demographically skewed, with underrepresentation of BIPOC, women, and older individuals~\cite{EEOC2024}. This impacts ethical priorities, as women and Black individuals prioritize responsible AI more than others~\cite{jakesch2022different}, while those from India see AI as authoritative and aspirational~\cite{kapania2022because}. Financial precarity among lower-income practitioners further limits engagement with ethical concerns~\cite{widder2023s, brophy2006system}. This demographic imbalance has a significant implication: AI practitioners place less value on responsible AI compared to the general public~\cite{jakesch2022different}. Broadening ethics studies to include diverse tech workers can address these gaps~\cite{nedzhvetskaya2022role}. However, research in tech ethics (FACCT and AIES) on marginalized groups often lacks sufficient attention, necessitating deeper engagement~\cite{birhane2022forgotten}. Our research examines how demographic backgrounds and perspectives on ethical issues in products affect the integration of ethical principles.  Understanding practitioners' individual perspectives on the software they develop is key to fostering equitable and responsible technology development.

\section{Methodology}

\subsection{Survey Development and Design}
To ensure survey robustness, we employed a multi-phase validation process. The survey was first reviewed by thirteen members of our software-focused research group, diverse in gender, race, and nationality, who provided feedback on its content and structure. Input from nine external software developers further informed its applicability. A pilot study with Prolific respondents (\textit{n} = 150) assessed response quality and gathered participant feedback, leading to iterative refinements across four rounds of discussion between two authors. 

Survey scale choices and question phrasing followed Krosnick's questionnaire design guidelines~\cite{krosnick2018questionnaire}. Key aspects evaluated included alignment with research objectives, clarity and neutrality of language, suitability of response scales, question order to minimize participant fatigue, inclusion of attention-check questions, and clarity of introductory text. To further enhance clarity, each survey section included a brief description defining terminology and context. \textit{Ethical concern} was defined as any software practice, decision, or feature potentially harmful to society or users, with a hover-over explanation provided for the term where applicable.

To ensure respondents shared a consistent understanding of the term "ethical issue," we utilized a taxonomy developed by Tjikhoeri et al.~\cite{tjikhoeri2024best}. This taxonomy was provided in the survey’s introductory section to shape respondents' interpretation of ethical issues and guide their responses. Additionally, this framework was used to construct the answer options for \textbf{Q1.2} and \textbf{Q1.4} (see Table~\ref{table:questions}). While predefined categories from the framework were presented, respondents could also select an "Other" option to include additional concerns.

Our study employs Tjikhoeri et al.’s framework~\cite{tjikhoeri2024best}, which builds on Wright’s foundational work~\cite{wright2011framework} and incorporates real-world ethical concerns from user feedback. We selected this framework to ground the ethical issues in user concerns and extend beyond AI-specific topics. Unlike Western AI ethics frameworks critiqued by Birhane et al.~\cite{birhane2022forgotten} for treating ethical principles as universal and abstract—detached from historical and contextual realities—Tjikhoeri et al.’s framework balances abstract principles (e.g., transparency, accountability) with practical issues such as misinformation, censorship, and scams, derived from user complaints in app reviews across a diverse range of applications. This dual emphasis ensures the framework's relevance to broader software systems and enhances its ability to address systemic ethical challenges. 

It has also been successfully applied in prior research to categorize and analyze ethical issues in software~\cite{sorathiya2024towards, karaccam2024uncovering, olson2023along, olson2024crossingmarginsintersectionalusers, elias2025unethicalsoftwareuserperceptions}. One such study~\cite{elias2025unethicalsoftwareuserperceptions} revealed that algorithmic unfairness is deeply entangled with user manipulation, opaque data collection, and the erosion of fundamental rights like privacy, accountability, and transparency. These findings highlight the interdependent nature of ethical issues in software, challenging the notion that concerns can be neatly categorized or resolved in isolation. By foregrounding end-user experiences and concerns, the framework shifts ethical evaluation from abstract ideals toward user-centered accountability, offering a novel and grounded lens for examining ethics in software development.

For the complete list of ethical issues and definitions, see Table~\ref{table:taxonomy} in the Appendix.

The survey comprises one informed consent question, mandatory questions on software experience, optional demographic questions (e.g., sex, gender, nationality, SES, race, ability, neurodivergence), and sections addressing ethical issues in software development \textbf{(RQ1)}, factors affecting treatment of ethical issues \textbf{(RQ2)} and developers' handling of such issues \textbf{(RQ3)} (see Table~\ref{table:questions} ). Most questions were multiple choice, with two multi-select (\textbf{(Q1.2)}, \textbf{(Q1.4)}) and one open-text (\textbf{(Q1.3)}). Anonymity was prioritized due to the likely prevalence of non-disclosure agreements (NDAs) among participants; thus, most questions were optional, allowing participants to avoid responses that could pose legal or personal risks. The complete survey is detailed in the replication package and was approved by the university's ethics committee. Informed consent was obtained from all participants.

\subsection{Participant Recruitment}
We recruited participants through Prolific\footnote{\url{https://www.prolific.com/}} using screening criteria to ensure participants had relevant experience in the software industry and to target specific demographic populations. Two rounds of recruitment were conducted. In the first round, no demographic filters were applied beyond filtering for software experience. This round ran on August 13, 2024, with participants compensated at an average rate of ~11 euros per hour for a median completion time of 9 minutes and 12 seconds. A total of 150 submissions were approved, with 24 rejected, 42 returned, and 6 timed out.

We conducted the second round on January 6, 2025, targeting a more balanced sample, focusing on participants from the Global South, female individuals, and BIPOC respondents. This round also included a software experience filter, with participants compensated at an average rate of ~11 euros per hour for a median completion time of 17 minutes and 39 seconds. A total of 68 submissions were approved, 14 were rejected, 17 returned, and 1 timed out.

To maintain data quality and mitigate the impact of financial incentives, three attention check questions were included in each round (see Section~\ref{app:attention_check}). These questions ensured participant attentiveness and validated self-reported software experience through basic technical knowledge assessments. Responses were rejected if participants failed any of the attention check questions and returned if participants started the survey then opted out.
The survey gathered demographic data from respondents, 43\% of whom identified as male, 41\% as female, 1\% as intersex, and 15\% preferred not to say. Regarding race, 33\% were White, 38\% Black, 4\% each East and South Asian, 3\% Mixed Race, and smaller proportions Latinx, Indigenous, and Middle Eastern, with 15\% not disclosing. Generationally, 40\% were Millennials, 27\% Gen Z, 15\% Gen X, 4\% Baby Boomers, and 14\% preferred not to say. Roles included software developers (37\%), data scientists (14\%), testers (12\%), project managers (11\%), consultants (8\%), and others. Experience levels ranged from 6\% with under 2 years to 10\% with over 20 years, with 38\% having 2–5 years and 20\% 6–10 years; 14\% did not disclose. Geographically, 33.8\% resided in Africa, 25.9\% in Western Europe, 16.7\% in North America, 6.5\% in Eastern Europe, and smaller percentages in South Asia, the Middle East, and Oceania, with 14.4\% not disclosing their location. Nationalities were similarly diverse, with 36.1\% identifying as African, 20.4\% as Western European, 13.0\% as North American, 7.4\% as Eastern European, and smaller percentages from South Asia (3.2\%), the Middle East (1.9\%), South America (0.9\%), Southeast Asia (0.5\%), East Asia (1.4\%), and Oceania (0.5\%), with 14.8\% preferring not to disclose. Our sample includes respondents from over 100 different companies, from a range of domains and sizes. For a full list, please refer to our replication package. Due to the demographic composition of software practitioners and the Prolific user base, our sample predominantly includes white men from the Global North (28.6\% , 62/217) and Black women from the Global South (28.1\% , 61/217) (see Appendix~\ref{app:int}).

\begin{table*}[htbp]
\small

\begin{tabular}{m{.2cm} m{14cm}}
\hline
 \textbf{RQ}  & \textbf{Question Text} \\ \hline

 RQ1 &  \textbf{Q1.1:} How would you describe the presence of ethical issues in the software developed by {[}Employer{]}?  \\ 

   &  \textbf{Q1.2:} Which ethical issue(s) are present in software developed by {[}Employer{]}?   \\ 

   &\textbf{Q1.3:} Are there any cases of ethical issues you would like to report in more detail? Please remember that this survey will remain entirely anonymous. Please detail the ethical issue(s) here:    \\

  &  \textbf{Q1.4:} In your experience at {[}Employer{]}, what types of ethical issue(s) do end-users report to {[}Employer{]}?  \\ 
  &  \textbf{Q1.5:} At {[}Employer{]}, were you asked to contribute to software you considered unethical? \\ \hline

 RQ2   & \textbf{Q2.1:} Were ethical issues discussed during your onboarding process at {[}Employer{]}? Consider any training sessions, materials, or discussions during your initial months at the company.   \\ 

    & \textbf{Q2.2:} Do employees' political views affect how they prioritise ethical issues in software developed by {[}Employer{]}? \\ 

    & \textbf{Q2.3:} Do employees' identities (e.g., race, ability, sexuality, gender, etc) affect how they prioritize ethical issues in software developed by {[}Employer{]}?  \\

    &  \textbf{Q2.4:} When developing new features at {[}Employer{]}, is it considered how they might affect certain demographic groups of users differently?   \\ 

    & \textbf{Q2.5:} When developing new features at {[}Employer{]}, which of the following factors is given the most consideration during decision-making?  \\ 

     &  \textbf{Q2.5a:} In your experience at {[}Employer{]}, how did considering [Q2.5\_answer] affect the way software teams considered their software's ethical issues?   \\ 

   & \textbf{Q2.6:} When writing code at {[}Employer{]}, is it clear how your code will fit into the overall product?  \\ 

   & \textbf{Q2.7:} When writing code at {[}Employer{]}, is it clear how your code will affect end-users?  \\ \hline

 RQ3 & \textbf{Q3.1:} At {[}Employer{]}, do you feel like your opinions on ethical issues regarding their software have been valued?  \\ 

   &  \textbf{Q3.2:} When you noticed ethical issues in {[}Employer{]}'s software, did you have the ability to intervene? \\ 

   & \textbf{Q3.3:} Did you give your perspective(s) to higher-level co-workers on ethical issues regarding the software developed by {[}Employer{]}? \\ 

  & \textbf{Q3.3a:} How often did sharing your perspective with higher-level co-workers on ethical issues influence the final product at {[}Employer{]}? \\ 
   & \textbf{Q3.4:} Were you comfortable sharing information about the ethical issues you've noticed at {[}Employer{]}?  \\ \hline






\end{tabular}
\caption{Outline of the survey questions used.}
\label{table:questions}
\end{table*}

\subsection{Survey Analysis}
The data analysis involved multiple statistical tests and data manipulation techniques to examine associations between variables from the survey dataset. We filtered all responses to remove missing or irrelevant responses before conducting the tests.

\subsubsection{Statistical Tests}
When analyzing ordinal versus categorical variables, we used the Kruskal-Wallis (H test)~\cite{kruskal1952use} and Mann-Whitney U tests~\cite{mann1947test}, while for ordinal versus ordinal relationships, we applied Spearman's rank correlation (SRC)~\cite{spearman2010proof}. These tests assessed the significance of differences in responses to \textbf{(Q1.1)} (ethical issue presence) and demographic variables across all Q-related questions. We applied the Mann-Whitney U Test when predictor variables had exactly two levels. This test compared differences in \textbf{(Q1.1)} and demographic characteristics responses between two groups. We used the Kruskal-Wallis Test when predictor variables had more than two levels (e.g., age groups). For statistically significant results under the Kruskal-Wallis Test, we ran a Dunn's test~\cite{dunn1964multiple} with a Bonferroni correction~\cite{weisstein2004bonferroni} to specify which groups were statistically significant from one another. We also performed tests with survey questions with a direct relationship: \textbf{(Q2.5)} and \textbf{(Q2.5a)} (Kruskal-Wallis then Dunn's) as well as \textbf{(Q3.3)} and \textbf{(Q3.3a)} (SRC). 

 To account for interactions between demographic variables, we implemented an iterative approach to test pairwise interactions using Kruskal-Wallis tests and post hoc Dunn's tests for significant results as well as calculating Cramer's V to test collinearity~\cite{cramer1999mathematical}, while ensuring robustness by filtering out groups with insufficient data points (<5). We employed Fisher's Exact Test~\cite{fisher1934statistical} to analyze contingency tables for binary categorical variables. For instance, we used a Fisher's Exact Test to examine the differences in frequencies of ethical issues types reported by practitioners \textbf{(Q1.2)} and those they perceive end-users to have \textbf{(Q1.4)}. 
\subsubsection{Open Question Analysis}
We use Saldana's \textit{Protocol Coding} approach~\cite{saldana2021coding} to analyze the open-text responses to \textbf{(Q1.3)}, which allowed respondents to explain their \textbf{(Q1.2)} responses. Protocol Coding follows a deductive approach. For \textbf{(Q1.2)}, we provide respondents with Tjikhoeri et al.'s~\cite{tjikhoeri2024best} taxonomy of ethical concerns as the response options. For \textbf{(Q1.3)}, we use the same taxonomy to classify open-text explanations, ensuring alignment between the response options and their elaborations.

\section{Results}
\subsection{Ethical Concerns of Practitioners and Users (RQ1)}
Overall, \textbf{(Q1.1)}, 56.68\% (123/217) of the respondents indicated that the software developed by their organization raised ethical concerns to varying extents (Several ethical issues = 8.29\% (18/217)), while 43.31\% (94/217) did not report any concerns (No ethical issues = 40.55\% (88/217), and Don't know = 2.76\% (6/217)). Statistically significant differences were observed for experience ($H = 9.72$, $p < 0.05$, H test), socioeconomic status ($H = 6.50$, $p < 0.05$, H test), nationality ($U = 4948.5$, $p < 0.01$, Mann-Whitney U), residence ($U = 4934$, $p < 0.01$, Mann-Whitney U), sex ($H = 9.81$, $p < 0.01$, H test), and gender ($U = 4922$, $p < 0.01$, Mann-Whitney U).

We found no statistically significant differences regarding the frequency of ethical concerns practitioners perceive \textbf{(Q1.2)} and those they perceive users to have \textbf{(Q1.4)}. Finally, 37.57\% (68/181) of respondents reported that they were asked to contribute to software they considered unethical \textbf{(Q1.5)}. However, the majority (62.43\%, 113/181) indicated that this was ``Never" the case. 
\subsubsection{Description of Ethical Issues.}
 We separate descriptions of ethical issues \textbf{(Q1.3)} by category (selected in \textbf{(Q1.2)}). This section details the types of ethical issues practitioners reported, offering insight into the ethical issues discussed in the rest of the survey questions and providing insight into how practitioners understood ethical issues given our definitions and framework. We do not consider them generalizable. In total, about half of respondents provided details on the ethical issues at their company. Responses were separated into 164 categories (a response could belong to multiple categories). We detail the ethical issues which have more than five responses: 

\textbf{Privacy.}
A fourth (42/164) of responses noted privacy concerns. Five respondents expressed concerns about user data being shared or sold to third parties without user consent. Additionally, eight respondents noted that companies were encouraging teams to gather more user data than disclosed, often leading to privacy concerns about how this data would be sold or used, highlighting that users were often not fully aware of this process. The lack of transparency in data practices was a key concern. Twelve respondents raised concerns about the handling of sensitive information, such as personal health data or financial information, and the risk of data breaches. Ten respondents mentioned concerns about the extensive collection and storage of user data. Companies were cited as having practices that collect significant amounts of data, with risks to user privacy. Issues like cookies, surveillance, and the lack of balance between data collection and privacy protection were prominent. Eight respondents mentioned general privacy concerns. One respondent mentioned challenges in keeping up with data protection laws. 

\textbf{Accessibility.} Accessibility issues were a recurring theme, with 12\% (19/164) of responses indicating that software development often neglects users with disabilities. Some noted that their applications lacked proper testing for accessibility features, such as compatibility with screen readers or adherence to color contrast standards. In two cases, respondents reported that accessibility concerns were deprioritized in favor of speeding up development. In one case, teams overlooked accessibility due to budget constraints. Two respondents reported that accessibility features were considered an afterthought rather than a core design element, another noted that Web Content Accessibility Guidelines (WCAG) are only a suggestion and not a requirement. Several respondents highlighted that inaccessible software design can exacerbate social inequalities and the digital divide, particularly for users with disabilities or limited technological literacy. Finally, participants noted difficulties users face in utilizing spatial data and mapping technologies and extra difficulties in protecting the data of  users with disabilities.

\textbf{Accountability.} Overall, 11\% (18/164) of responses raised accountability as a concern. One respondent highlighted the lack of accountability in their company for incorrectly judging test cases. Similarly, another described inadequate testing of updates, which led to severe bugs and vulnerability remediations that failed to address the correct issues. Another practitioner mentioned ongoing debates about who should be held responsible when ethical issues arise. Three respondents pointed specifically to management mishandling ethical concerns. One noted that budget constraints caused senior management to overlook these issues, multiple respondents mentioned that higher-ups were not held accountable for mistakes, and others expressed concerns about a feature allowing admins to access client passwords. Nepotism was also mentioned. Three respondents noted that their employers were developing software for countries known for mistreating minorities, and one mentioned unfair labor practices. Additionally, one practitioner pointed out that feature development sometimes conflicts with end-user interests, with corporate priorities taking precedence. Another admitted that many back-end processes often skirt around ethical concerns.

\textbf{Transparency.}
Eight percent (14/164) of responses detailed transparency issues. Seven respondents expressed concerns related to the misrepresentation of products. These respondents noted the frequent use of corporate buzzwords in both internal and external communication, making it difficult for users and customers to fully understand the products. According to one practitioner, proprietary software and hardware made it challenging for users to grasp how their data was being used. In some cases, respondents reported that requirements were not fully conveyed to customers, and there was deliberate downplaying of product issues upon request from clients. One respondent raised concerns about the lack of transparency in terms of auditing practices, noting that there were insufficient audits to ensure transparency in their company’s processes. One respondent mentioned a lack of transparency specifically related to algorithms, highlighting that users and customers were not informed about how algorithms functioned. Another participant indicated a lack of transparency within teams. Four respondents brought up general transparency concerns, providing no detail.

\textbf{Environmental Issues.}
Ten respondents raised concerns about the environmental footprint of software development, including the energy consumption of software products and the production, distribution, and disposal of hardware. One respondent pointed out that Green IT recommendations were not taken seriously by their company.


\begin{figure}
\includegraphics[width=\columnwidth]{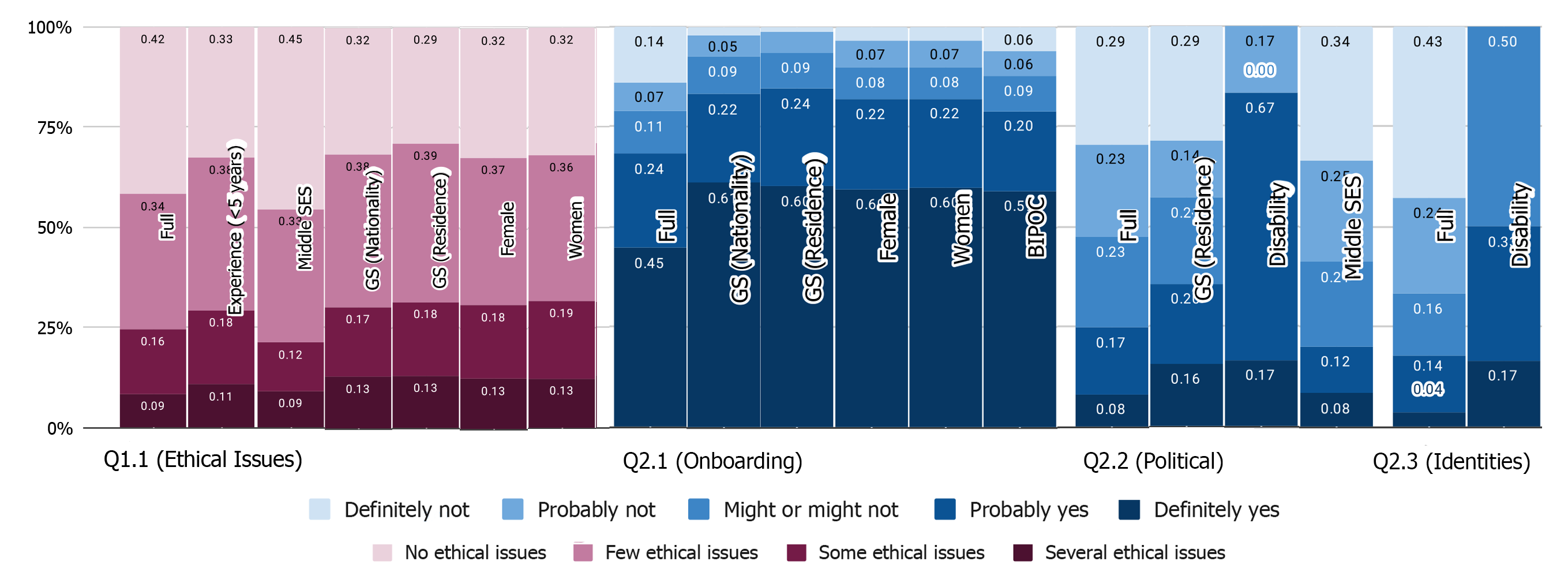}
  \caption{Responses to research questions Q1.1 through Q2.3, with the \textit{Full} sample, then stratified by marginalized demographic groups, including BIPOC, Global South (based on nationality and residence), women (gender), individuals with disabilities, socio-economic status (middle class), participants with less than 5 years of experience, and those identifying as female. The demographic groups presented were selected based on statistically significant results. The x-axis represents the research sub-questions, while the y-axis shows the percentage distribution of responses across Likert categories.}
  \label{fig:rq1-2}
\end{figure}

\subsection{ Ethical Concerns and Software Processes \textbf{(RQ2)}}
A notable number of respondents, 13.82\% (30/217), reported that ethical issues were ``Definitely not'' discussed during the onboarding process, while 44.70\% (97/217) reported that they ``Definitely were'' \textbf{(Q2.1)} (see Figure~\ref{fig:rq1-2}). There was a statistically significant difference in responses about the onboarding process for nationality (Mann-Whitney U, $U=5710.5$, $p < 0.001$), residence ($U=5390$, $p < 0.001$), gender ($U=5429$, $p < 0.001$), sex (H test, $H=12.04$, $p < 0.01$), and race (H test, $H=27.55$, $p < 0.001$). A Dunn's test revealed that women reported significantly more discussions about ethical issues during onboarding compared to men ((Dunn's test, $Z=-3.34$, $p=0.0025$). Additionally, Black respondents reported significantly more discussions compared to White respondents ((Dunn's test, $Z=-4.35$, $p<0.001$).

When asked whether employees' political views affect how they prioritize ethical issues \textbf{(Q2.2)}, 29.28\% (53/181) of respondents indicated ``Definitely not,'' while 22.65\% (41/181) selected ``Might or might not,'' and 8.29\% (15/181) selected ``Definitely yes'' (see Figure~\ref{fig:rq1-2}). A statistically significant positive correlation was observed between reported ethical issue presence and the perception that employees' political views influence prioritization \textbf{(Q2.2)} (Spearman's $\rho=0.409$, $p<0.001$).  There was a statistically significant difference in responses by residence (Mann-Whitney U test, $U=3608.5$, $p < 0.05$), ability (Mann-Whitney U test, $U=683.5$, $p < 0.05$), and socioeconomic class (H test, $H=7.08$, $p < 0.05$). Those from the middle class reported significantly more influence of political views compared to those from the upper class (Dunn's test, $Z=2.66$, $p<0.05$).

In general, participants agreed less with the idea that employees' identities (e.g., race, ability, sexuality, gender) affect the prioritization of ethical issues \textbf{(Q2.3)}; 42.78\% (77/180) indicated ``Definitely not,'' while 15.56\% (28/180) selected ``Might or might not,'' and 3.89\% (7/180) selected ``Definitely yes'' (see Figure~\ref{fig:rq1-2}). There is a significant positive correlation between ethical issues and employees' identities affecting ethical consideration \textbf{(Q2.3)} (Spearman's $\rho=0.306$, $p<0.001$). Additionally, there was a statistically significant difference in responses by ability (Mann-Whitney U test, $U=724.5$, $p<0.01$).

\begin{figure}
\includegraphics[width=\columnwidth]{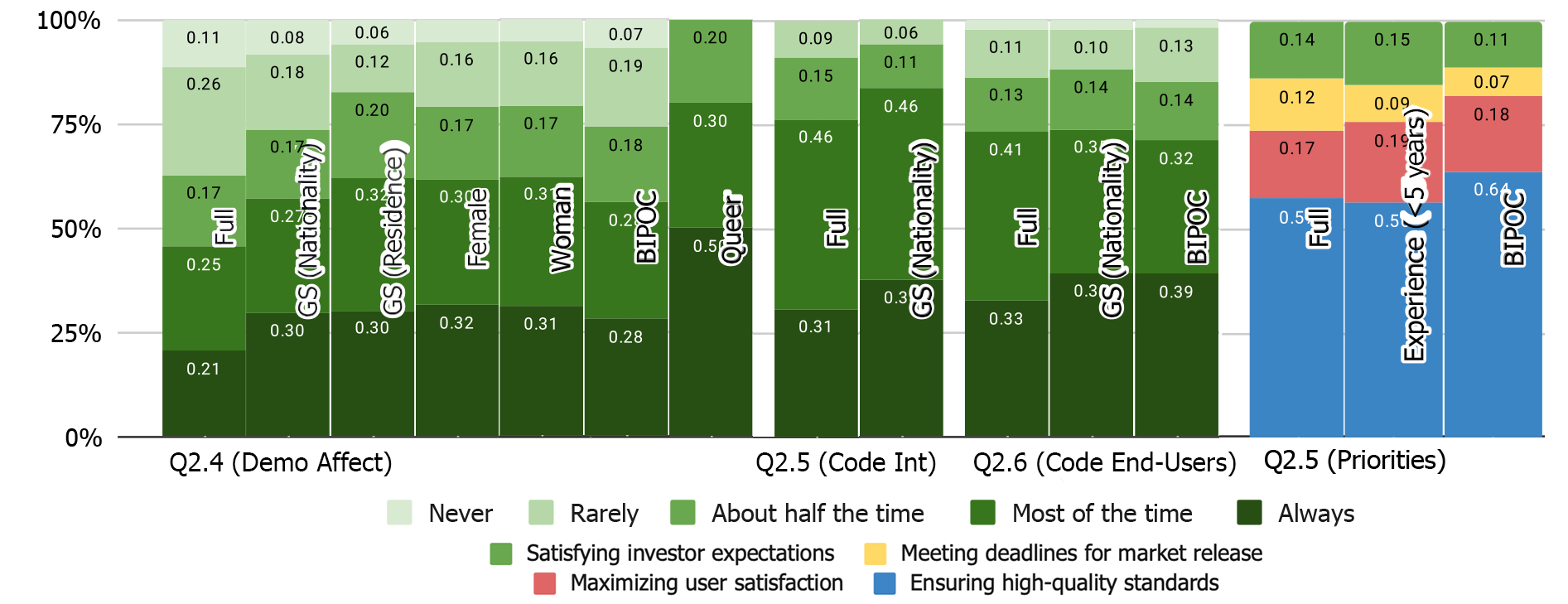}
  \caption{Responses to research question Q2.4 to Q2.7, with the \textit{Full} sample, then stratified by demographic groups, including BIPOC, Global South (nationality and residence), women, and individuals with disabilities. The x-axis represents the demographic stratifications, while the y-axis displays the percentage distribution of responses across Likert categories.}
  \label{fig:rq2.2}
\end{figure}

Nearly half of respondents indicated that demographic considerations are often neglected when developing new features \textbf{(Q2.4)}, with 37.22\% (67/180) selecting ``Rarely'' or ``Never'' (see Figure~\ref{fig:rq2.2}). There were significant differences in responses by nationality (Mann Whitney U test, $U=2494$, $p<0.01$), residence ($U=2678$, $p<0.001$), gender ($U=2660$, $p<0.001$), sex (H test, $H=13.80$, $p<0.05$), race ($H=22.82$, $p<0.01$), and sexuality ($U=213.5$, $p<0.01$). Additionally, women were significantly more likely to report that demographic considerations were made more compared to men (Dunn's test, $Z=-3.71$, $p<001$), while Black respondents report that demographic considerations were made more compared to White respondents (Dunn's test, $Z=-3.65$, $p<0.01$).

When asked which factors are most prioritized during the development of new features \textbf{(Q2.5)}, 57.22\% (103/180) of respondents indicated that ensuring high-quality standards (e.g., performance, safety, accessibility) is the top priority, while meeting deadlines for market release (12.22\%, 22/180), satisfying investor expectations (13.89\%, 25/180), and maximizing user satisfaction (16.67\%, 30/180) were less frequently selected (see Figure~\ref{fig:rq2.2}). A Kruskal-Wallis test revealed a statistically significant difference in ethical issue reporting based on the prioritization of these factors \textbf{(Q2.5)} ($\chi^2 = 19.13$, $p < 0.001$).  Similarly, those prioritizing "Satisfying investor expectations" reported significantly \textit{more} ethical issues than those emphasizing "Ensuring high-quality standards" ($Z = 2.83$, $p < 0.05$). Dunn's test indicated that respondents who prioritized "Meeting deadlines for market release" reported almost significantly \textit{more} ethical issues compared to those prioritizing "Ensuring high-quality standards" ($Z = 3.59$, $p = 0.08$). Further analysis revealed significant differences in prioritization by experience (Kruskall Wallis, $H=10.63$, $p<0.05$) and race ($\chi^2=12.17$, $p<0.05$).

When examining whether software priorities interfered with or enhanced ethical considerations \textbf{(Q2.5a)}, 64\% (14/22) of respondents indicated interference when prioritizing meeting deadlines, while 56\% (14/25) indicated interference for satisfying investor expectations. For \textbf{(Q2.5a)}, a Kruskal-Wallis test revealed a significant difference in the response distribution ($\chi^2 = 19.13$, $p < 0.001$). Dunn's test further indicated that respondents prioritizing "Meeting deadlines for market release" or "Satisfying investor expectations" experienced significantly \textit{more} interference compared to those prioritizing "Ensuring high-quality standards" ($Z = 3.59$, $p < 0.01$ and $Z = 2.83$, $p < 0.05$, respectively). Similarly, respondents prioritizing "Meeting deadlines for market release" reported significantly more interference compared to those prioritizing "Maximizing user satisfaction" ($Z = 3.03$, $p < 0.05$).

Regarding clarity of code integration \textbf{(Q2.6)}, 8.99\% (16/181) of respondents reported that they understood how their code fits into the product ``Rarely'' or ``Never,'' while 76.41\% (136/181) reported clarity ``Always'' or ``Most of the time'' (see Figure~\ref{fig:rq2.2}).  Mann-Whitney U test revealed a statistically significant difference in clarity of code integration by nationality ($U=2484.5$, $p<0.05$).

For clarity on how code affects end-users \textbf{(Q2.7)}, 32.57\% (57/181) of respondents reported ``Always,'' while 13.14\% (23/181) reported ``About half the time'' (see Figure~\ref{fig:rq2.2}). A Mann-Whitney U test revealed a statistically significant difference in clarity by nationality ($U=2478.5$, $p<0.05$), and a Kruskal-Wallis test showed significant differences by race ($\chi^2=12.82$, $p<0.05$). 
\begin{figure}
\includegraphics[width=\columnwidth]{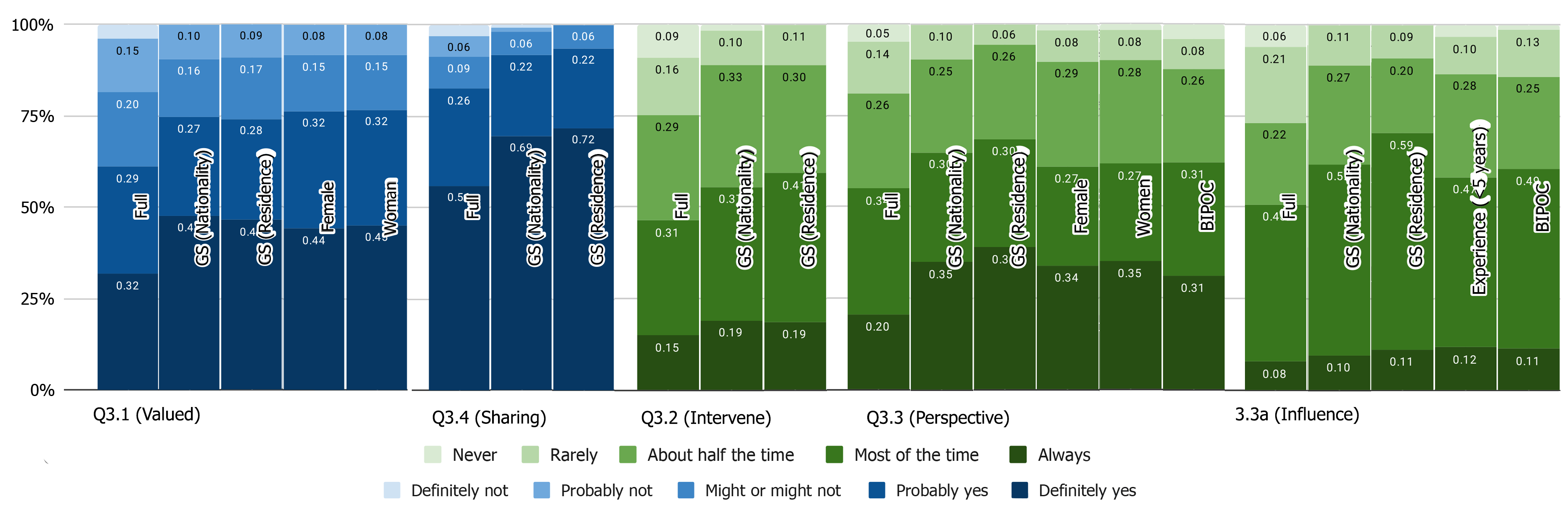}
  \caption{Responses to research questions Q3.1 through Q3.3a, with the \textit{Full} sample, then stratified by demographic groups, including BIPOC, Global South (nationality and residence), women, participants with less than 5 years of experience, and those identifying as queer. The demographic groups presented were selected based on statistically significant results. The x-axis represents the research sub-questions, while the y-axis shows the percentage distribution of responses across Likert categories.}
  \label{fig:rq3}
\end{figure}
\subsection{Ethical Concerns' Mitigation \textbf{(RQ3)}}
The first five questions detailed in this section only appeared to participants who indicated that there were ethical issues in their software (123 total participants). Only 60.98\% (75/123) feel that their opinions on ethical issues are valued \textbf{(Q3.1)} (see Figure ~\ref{fig:rq3}). Participants' perspectives on their opinions being valued were significantly associated with nationality (Mann-Whitney U test, $U=1878$, $p<0.001$), residence ($U=1824$, $p<0.01$), gender ($U=1820$, $p<0.001$), and sex (Kruskal-Wallis, $\chi^2=13.13$, $p<0.01$). Dunn's post hoc tests revealed that women were significantly more likely to feel their opinions on ethical issues were valued compared to men (Dunn's test, $Z=-3.43$, $p<0.01$), while Black respondents were significantly more likely to feel this way compared to White respondents (Dunn's test, $Z=-3.65$, $p<0.01$).

Regarding whether practitioners feel they can intervene when ethical issues arise \textbf{(Q3.2)}, 24.79\% (30/121) of respondents indicated "Rarely" or "Never" (see Figure ~\ref{fig:rq3}). Nationality ($U=975.5$, $p<0.01$) and residence ($U=956$, $p<0.01$) were significantly associated with responses to this question.

Sharing perspectives with higher-level co-workers yielded mixed results \textbf{(Q3.3)}, with 18.85\% (23/122) of respondents indicating that they gave their input "Rarely" or "Never" on ethical issues (see Figure ~\ref{fig:rq3}). Responses significantly differed by nationality ($U=986$, $p<0.001$), residence ($U=1098$, $p<0.001$), gender ($U=954.5$, $p<0.001$), sex (Kruskal-Wallis, $\chi^2=11.04$, $p<0.01$), and race (Kruskal-Wallis, $\chi^2=23.28$, $p<0.01$). Dunn's test showed that women were significantly more likely than men to share perspectives with higher-level co-workers on ethical issues (Dunn's test, $Z=-3.08$, $p<0.01$).

When asked how often these perspectives influenced the final product \textbf{(Q3.3a)}, 20.69\% (24/116) of the respondents answered "Rarely," and 6.03\% (7/116) said "Never" (see Figure ~\ref{fig:rq3}). Responses significantly varied by years of experience (Kruskal-Wallis, $\chi^2=12.82$, $p<0.05$), nationality ($U=1107$, $p<0.001$), residence ($U=1161$, $p<0.001$), and race (Kruskal-Wallis, $\chi^2=27.18$, $p<0.001$). Dunn's test revealed that Black respondents (Dunn's test, $Z=-3.84$, $p<0.01$) and South Asian (Dunn's test, $Z=-3.13$, $p<0.05$) respondents were significantly more likely than White respondents to report that their feedback influenced the final product.

We also found a statistically significant positive correlation between participants who gave input to higher-level colleagues frequently and those who said their feedback often influenced the final product (Spearman's $\rho=0.585$, $p<0.001$). Lastly, there is also a clear indication that respondents are generally comfortable sharing information about ethical issues for the survey \textbf{(Q3.4)}, with 82.41\% (178/216) of the respondents answering "Definitely yes" or "Probably yes." Participants' comfort levels when sharing ethical concerns about their employers varied significantly between nationalities (Mann-Whitney U test, $U=5289$, $p<0.01$) and residential backgrounds ($U=5250$, $p<0.01$) \textbf{(Q3.4)}.


\subsection{Interactions between Demographic Characteristics}
We examine the interaction between demographic variables, driven by two key factors: (1) the sampling imbalance, with a skew toward two groups—white men from the Global North and Black women from the Global South (see Appendix~\ref{app:int})—and (2) the complex and intrinsic interrelation among demographic characteristics. Our results reveal particularly high collinearity (Cramér's V, >.8) between race, residence, and nationality, as well as between gender and sex, and gender and nationality, race, or residence. Notably, we observed moderate to high collinearity (Cramér's V, >.3) among nearly all demographic variables, especially residence—a finding that aligns with the significant influence of residential context on various aspects of life (see Fig.~\ref{fig:collinearity_heatmap} in Appendix~\ref{app:correlation}). Consequently, we conducted additional Kruskal-Wallis tests and Dunn’s post hoc analyses to explore intersectional interactions (see Appendix~\ref{app:correlation}). On the Kruskal Wallis interaction tests, significant differences emerged between men from the Global North and women from the Global South across multiple survey questions, including \textbf{(Q1.1)}, \textbf{(Q2.1)}, \textbf{(Q2.4)}, \textbf{(Q3.1)}, \textbf{(Q3.2)}, \textbf{(Q3.3)}, and \textbf{(Q3.3a)} (often for both residence and nationality; see Appendix~\ref{app:correlation}). Additionally, significant differences were identified between women from the Global South and the Global North for \textbf{(Q2.2)} and \textbf{(Q3.3a)}. For \textbf{(Q3.3a)}, differences were also observed between women from the Global North and men from the Global South. These findings highlight the importance of understanding demographic characteristics as deeply intersectional and interwoven and provide context for interpreting the results for all questions.
\section{Discussion} 
Our results demonstrate that (\textbf{\textit{RQ1}}) marginalized practitioners report greater ethical sensitivity to ethical issues; (\textbf{\textit{RQ2}}) factors such as organizational norms and individual demographics significantly influence ethical integration; and (\textbf{\textit{RQ3}}) many practitioners feel unable to mitigate ethical issues, although marginalized practitioners report a greater ability to do so.

\subsection{Ethical Concerns Identified by Practitioners and Attributed to Users (RQ1) }
A slight majority (57\%) of software practitioners reported ethical concerns. Among those who reported ethical concerns, practitioners from marginalized groups (e.g., women, BIPOC, disabled individuals) tended to report them at a higher frequency. Statistically significant differences were observed for several demographic factors, including socioeconomic status, years of experience, nationality, residence, sex, and gender. These findings align with prior work emphasizing the importance of diverse team composition in uncovering ethical risks~\cite{pant2024ethics, widder2023s, jakesch2022different}.

We found no statistically significant differences regarding the frequency of ethical concerns practitioners perceive \textbf{(Q1.2)} and those they believe users perceive \textbf{(Q1.4)}. However, this may point to a potential gap in developers' awareness of how users' ethical values might differ from their own. As previous research has shown that AI practitioners tend to undervalue responsible AI principles compared to the general public~\cite{jakesch2022different}, this gap indicates the need for proactive user feedback mechanisms that incorporate best practices, such as active employee participation, feedback collection and triangulation across multiple sources (e.g., social media, user metrics, and conventional feedback), and prompt responses to user concerns~\cite{li_unveiling2024}. 
\subsection{Factors Influencing Ethics Integration in Software Development (RQ2)} A substantial number of practitioners (21\%) reported receiving minimal or no training on ethical issues during their initial months in their roles. Notably, our findings suggest that the delivery and quality of onboarding training may vary across demographic groups. This discrepancy is particularly concerning given the significant responsibility placed on practitioners to address ethical challenges~\cite{widder2023s}, highlighting the need for systematic approaches, such as legislation mandating ethics training for software practitioners, modeled after health privacy training requirements (§164.530(b)(1) of the HIPAA Privacy Rule~\cite{hipaa_training}), to ensure universally implemented and culturally sensitive onboarding.
Additionally, research shows that `privacy champions' consider code reviews and practical training sessions to have a greater impact on privacy end-results than on-boarding training~\cite{10.1145/3411764.3445768}. Thus, future research should also investigate the treatment of ethical concerns during code reviews and training sessions. 

Furthermore, respondents who indicated that employees' political views and identities influenced practitioners' prioritization of ethical issues also reported significantly more ethical issues compared to those who indicated no such influence. This finding could indicate that respondents perceive practitioners' political views and identities to have a negative impact on ethical consideration, potentially biasing the process based on their personal background or could point to greater ethical sensitivity among those who recognize the relationship between background and ethical prioritization. Thinking ethical integration should be neutral aligns with a dominant culture in the tech industry—\textit{the myth of neutrality}, where technologies are often perceived as value-neutral tools determined solely by user choice, ignoring the significant influence of social, cultural, and political constraints~\cite{mowshowitz1984computers}. Assuming neutrality distorts the analysis of ethical considerations and reinforces a reactive, uncritical approach to addressing the societal implications of technology. 

In conjunction, a majority of respondents (67\%) did not believe that practitioners' identities influence the prioritization of ethical issues during development. This could indicate a lack of awareness regarding how demographic backgrounds shape ethical perspectives~\cite{fumagalli2010gender, grgic2018human}. This finding is especially critical when considering the demographic imbalance in software development roles~\cite{EEOC2024}, meaning that perspectives from underrepresented groups might be systematically undervalued and disregarded in groups where this ideology prevails. Research already shows that minority groups are less likely to speak up in majority dominant groups~\cite{wuestenenk2025influence}.

Disabled individuals were significantly more likely to indicate that identity and political affiliation influence ethical considerations. This group likely brings critical perspectives on integrating ethics, particularly in relation to accessibility and discrimination. However, to the authors' knowledge, research within the FACCT community has only focused on algorithmic bias affecting disabled individuals~\cite{givens_dis_2020, buyl2022tackling, gadiraju2023wouldn, glazko2024identifying}. We therefore advocate for expanded research into the pivotal role disabled individuals play in advancing fairness and equity in technology.

Additionally, some respondents reported that demographic considerations are neglected during feature development. As one participant noted, it seems that these considerations tend to be prioritized only when there is a legal directive rather than a guideline or standard. Interestingly, marginalized populations often reported demographic considerations being made more frequently. 

Our analysis of software development priorities revealed statistically significant differences, indicating that teams prioritizing high-quality standards are more effective at addressing ethical issues compared to those emphasizing deadlines or investor expectations. This finding offers broader empirical support across software practitioners, providing support that market-driven pressures can hinder the consideration of ethics as suggested by prior interview studies in AI~\cite{ali2023walking, rakova2021responsible, madaio2020co, gray2019ethical} and software ethics~\cite{widder2023s}. 

Although we may not be able to restructure economic systems of power, we can make systematic changes to the processes which dictate how software is made. Our finding regarding meeting deadlines hindering ethical consideration is particularly relevant in the context of agile development, where rapid iterations may hamper comprehensive ethical evaluations. While agile practices are widely adopted, our results imply that slower, more deliberate development processes could improve ethical considerations. Future work could explore modifications to agile methods that incorporate ethical checkpoints, as suggested by Hussain et al.~\cite{hussain2022can} who propose introducing roles, artefacts, cultural practices and ceremonies focused on human values within agile frameworks. 

\subsection{Mitigating Ethical Concerns in Software Products (RQ3) } 
Many respondents (25\%) indicated that they did not feel empowered to intervene when they noticed ethical issues in their employer's software, while 19\% were uncertain about whether their ethical concerns would be considered by higher-level co-workers. These results are troubling when considering that handling ethical issues is highly dependent on practitioners' ability to convince decision-makers to allocate resources to that pursuit~\cite{widder2023s}. Previous research indicates that when employees' ethical concerns are dismissed, they often adopt various tactics, such as reframing their concerns to align with organizational norms or presenting them as financial or reputational risks to the employer~\cite{wong2021_tactics, widder2023s}, or alternatively, reducing their level of effort and engagement~\cite{widder2023s}.

Education and training are therefore necessary to empower practitioners to advocate for ethical design and inform them on the best tactics to navigate these situations. Previous research suggests that educators and trainers can prepare future software engineers to advocate for ethics in the workplace by incorporating strategies such as inviting guest speakers with real-world experience, facilitating hands-on activities like designing communication tools, and assigning projects that involve identifying ethical issues and engaging stakeholders~\cite{rivera2024teaching}. These methods help students build practical skills in framing ethical arguments, presenting complex concepts, and navigating workplace challenges effectively. 

Further, we find that there is a statistically significant positive correlation between how often employees share their ethical concerns with higher-level coworkers and those who consider their discussions to frequently impact the final product. This result supports the notion that there may be a negative feedback loop where, if employees' ethical concerns are disregarded, they may speak up less. These findings emphasize the importance of incorporating opportunities within education to help students develop and practice skills for engaging in effective ethical dialogues with those in decision-making roles.

Our findings indicate that marginalized groups report feeling more valued, are more likely to share their perspectives, intervene and influence ethical product development. Based on our intersectional analysis, the greatest difference appears between two main groups: white men from the Global North and Black women from the Global South. The latter group's increased capacity to handle ethical concerns is likely influenced by the representation of South African practitioners within this sample (n = 72). Academic literature widely recognizes Ubuntu as the dominant cultural philosophy in sub-Saharan Africa~\cite{mangaliso2018invoking}, a framework that asserts human identity is rooted in communal, moral responsibility. This cultural foundation aligns with our findings that nationality and residence significantly influenced whether respondents felt their ethical concerns were valued and whether they could intervene when ethical issues arose.

Research also shows that South Africa follows an inclusive rather than exclusive corporate governance regime~\cite{khomba2013shaping}. Unlike the shareholder-centric models prevalent in the US, South Africa’s corporate governance framework emphasizes broad stakeholder involvement, ensuring that decision-making processes account for not only shareholders but also employees, communities, and other affected parties. This inclusive governance structure aligns with Ubuntu principles, reinforcing collective responsibility and ethical deliberation as integral to professional environments. The structural role of inclusion is reflected in our results, where marginalized respondents more frequently gave input to higher-level colleagues and were also significantly more likely to see their feedback shape final product decisions.

Ubuntu shapes organizational culture in multiple ways, influencing communication styles, decision-making processes, and productivity approaches [4]. Unlike Western business models that prioritize efficiency and individual accountability, Ubuntu emphasizes face-to-face communication over written exchanges and fosters consensus-driven decision-making, ensuring dissenting voices are acknowledged rather than overridden by majority rule. Employees are encouraged to express their perspectives, and while this process can be time-intensive, it is valued for fostering cohesion and ethical responsibility. Additionally, Ubuntu-oriented work environments prioritize social harmony over rigid productivity metrics~\cite{mangaliso2018invoking}.

This emphasis on collective responsibility fosters broader accountability, where ethical concerns are addressed through dialogue rather than unilateral directives. The reliance on direct, participatory communication enhances transparency, ensuring that ethical considerations are collectively acknowledged and scrutinized rather than obscured by hierarchy. The significance of transparency and accountability was further reflected in our results, as Global South respondents were more comfortable sharing ethical concerns about their employers.

Madaio et al. found that AI fairness efforts are largely driven by individual advocates~\cite{madaio2020co}, reflecting the Western philosophical emphasis on individualism seen in utilitarianism and deontology. In contrast, Ubuntu shifts ethical responsibility to communitarianism~\cite{khomba2013shaping}, embedding fairness, accountability, and transparency into participatory decision-making rather than relying on individuals to push for ethical considerations within unsupportive institutional frameworks.

\section{Limitations}
Our study's sample size was constrained by our university's limited budget, which may affect the generalizability of the survey results, particularly given that most questions were optional. This decision introduced the potential for non-response bias. Nevertheless, allowing optional responses was necessary to ensure respondents’ anonymity and comfort, given the sensitive nature of the topic. The use of the Prolific platform for participant recruitment introduced a financial incentive, which may have led some respondents to prioritize speed over the quality of their responses. To mitigate this, we incorporated three attention check questions designed to ensure respondents were engaged and possessed the necessary software development knowledge. A total of 38 respondents were excluded for failing these attention checks, including one participant who completed the survey in just 19 seconds. 

Additionally, the survey may suffer from self-selection bias, as participation was voluntary. Respondents who chose to complete the survey may have stronger opinions or experiences with ethical concerns, potentially skewing the results. As with all self-reported data, the survey is susceptible to self-reporting bias. Respondents may have provided socially desirable answers, potentially leading to the underreporting of ethical concerns or an overly positive assessment of the societal impact of the software they develop. Additionally, imperfect recollection of events may have further affected the accuracy of responses. Given that many software practitioners are bound by non-disclosure agreements (NDAs), it is likely that some respondents were constrained from fully disclosing their experiences, particularly when detailing ethical concerns that could result in reputational or financial harm to their employer. The fear of potential job repercussions may have further limited the responses concerning ethical issues.
\section{Conclusion}
This study sheds light on the complex landscape of ethical considerations in software development, revealing critical gaps and opportunities for improvement. Our findings demonstrate the significant influence of demographic factors on practitioners’ ethical sensitivity and empowerment, with marginalized groups often reporting heightened awareness and capacity to address ethical challenges. However, systemic barriers, such as organizational norms prioritizing market pressures over ethical principles and the myth of technological neutrality, persist in hindering effective ethical integration. By broadening the scope of ethical concerns to include user-generated issues and leveraging practical frameworks, this research underscores the importance of diverse perspectives and proactive interventions. To advance ethical practices in software development, reforms in education, onboarding, and development processes are essential, along with systematic changes to encourage inclusive, thoughtful, and responsible computing. Future work should focus on refining agile methodologies to incorporate ethical checkpoints, examining ethics education, and fostering environments where diverse voices are valued and supported in addressing ethical challenges.

\bibliographystyle{acm}
\bibliography{citations}
\section*{Appendix}
\appendix
\section{Methods}

\subsection{Attention Check Questions}
\label{app:attention_check}
\begin{enumerate}
    \item \textbf{Attention Check 1: Programming Language Selection} \\
    \textit{Software skills are critical for development at \texttt{\$\{e://Field/Company\}}. This is an attention check question. Please select the programming language.}
    \begin{itemize}
        \item Docker
        \item Javascript
        \item Scrum
        \item GUI Design
    \end{itemize}

    \item \textbf{Attention Check 2: Ethical Issues Severity} \\
    \textit{Some ethical issues regarding software developed by \texttt{\$\{e://Field/Company\}} are more severe than others. This is an attention check question. You must respond. (Please select Accessibility.)}
    \begin{itemize}
        \item Manipulation
        \item Dark Patterns
        \item Targeted Advertising
        \item Accessibility
        \item Surveillance
    \end{itemize}

    \item \textbf{Attention Check 3: Version Control System} \\
    \textit{There are many types of software systems. Please answer the term that is a version control system. This is an attention check. You must fill it out.}
    \begin{itemize}
        \item Hadoop
        \item Tensorflow
        \item Apache Spark
        \item Git
        \item Prolog
    \end{itemize}
\end{enumerate}

\begin{table*}[h!]
\centering
\small
\caption{Ethical Issues' Definitions}
\begin{tabular}{p{0.2\textwidth}|p{0.8\textwidth}}
\hline
\textbf{Ethical Issue}  & \textbf{Definition}  \\\hline

\rowcolor{lightgray}Accessibility & The application does not include people with special needs or disabilities. This concern can be about the usage of the application itself or about a service the application is offering.\\

Accountability & The user experienced an issue when using the application or its service. The user could not find the software company responsible for solving the issue.\\

\rowcolor{lightgray}Addiction & The user mentions how they are addicted to the application or describes that they use it excessively.\\

Censorship & The application deliberately hides certain information, or certain users' content or profiles are deliberately removed or demoted.\\

\rowcolor{lightgray}Content Theft & Content from a user is stolen or used without permission from the original creator.\\

Cyberbullying & The platform's community is being harmful, abusive, or unhealthy by practicing hateful communication via the application.\\

\rowcolor{lightgray}Discrimination & The application user is being discriminated against by the application, its services, or its community. This concern also includes users who have an issue with having fewer functionalities available to them because they live in a different geographical area.\\

Sustainability & The user says something about the negative impact the application has on the environment.\\

\rowcolor{lightgray}Harmful Advertising & The user notices the presence of deceiving, misleading, or harmful advertisements throughout the application.\\

Identity Theft & Someone is using the identity of someone else on this application. This concern also applies to catfishing, creating fake profiles to trick and deceive other users on the platform. \\

\rowcolor{lightgray}Inappropriate Content & The application contains content other than advertisements that are disturbing to certain groups of people.\\

Privacy & The users' identity and data are not kept secure or used for purposes other than what the user gave consent to. This concern also includes when an account is hacked.\\

\rowcolor{lightgray}Safety & The usage of this app has led to health issues or other safety risks. This concern can be about the usage of the application itself or its services.\\

Scam & The user has been scammed or came into contact with a scammer through the application. This concern can occur through the application only or its services. A scammer deceives another to gain something, usually money or goods.\\

\rowcolor{lightgray}Misinformation & False information is spread through this application.\\

Transparency & The motives, risks, and implications are unclear to the user when using this application or a service the application provides.\\\hline

\end{tabular}
\label{table:taxonomy}
\end{table*}

\subsection{Intersectional Demographic Information}
\label{app:int}
\begin{table}[]
\caption{Intersectional Demographic Information}
\centering
\begin{tabular}{l|p{0.6cm}p{1cm}|p{0.6cm}p{0.6cm}p{0.6cm}p{0.6cm}p{.6cm}p{0.6cm}p{0.8cm}|p{0.6cm}p{0.6cm}p{0.6cm}p{0.6cm}}
\hline
 & \textbf{Man} & \textbf{Woman} & \textbf{White} & \textbf{Black} & \textbf{Latinx} & \textbf{East Asian} & \textbf{Indig.} & \textbf{South Asian} & \textbf{Middle Eastern} & \textbf{Global South (R)} & \textbf{Global North (R)} & \textbf{Global South (N)} & \textbf{Global North (N)} \\\hline
\textbf{Man} & 94 & 0 & 63 & 14 & 2 & 5 & 1 & 3 & 1 & 12 & 82 & 19 & 74 \\
\textbf{Woman} & 0 & 90 & 7 & 69 & 1 & 3 & 0 & 6 & 1 & 66 & 24 & 74 & 16 \\ \hline
\textbf{White} & 63 & 7 & 72 & 0 & 0 & 0 & 0 & 0 & 0 & 2 & 70 & 3 & 69 \\
\textbf{Black} & 14 & 69 & 0 & 83 & 0 & 0 & 0 & 0 & 0 & 68 & 15 & 72 & 10 \\
\textbf{Latinx} & 2 & 1 & 0 & 0 & 3 & 0 & 0 & 0 & 0 & 1 & 2 & 3 & 0 \\
\textbf{East Asian} & 5 & 3 & 0 & 0 & 0 & 8 & 0 & 0 & 0 & 0 & 8 & 3 & 5 \\
\textbf{Indigenous} & 1 & 0 & 0 & 0 & 0 & 0 & 1 & 0 & 0 & 0 & 1 & 0 & 1 \\
\textbf{South Asian} & 3 & 6 & 0 & 0 & 0 & 0 & 0 & 9 & 0 & 4 & 5 & 7 & 2 \\
\textbf{Middle Eastern} & 1 & 1 & 0 & 0 & 0 & 0 & 0 & 0 & 2 & 0 & 2 & 1 & 1 \\ \hline
\textbf{Global South (R)} & 12 & 66 & 2 & 68 & 1 & 0 & 0 & 4 & 0 & 78 & 0 & 76 & 2 \\
\textbf{Global North (R)} & 82 & 24 & 70 & 15 & 2 & 8 & 1 & 5 & 2 & 0 & 108 & 17 & 90 \\
\textbf{Global South (N)} & 19 & 74 & 3 & 72 & 3 & 3 & 0 & 7 & 1 & 76 & 17 & 93 & 0 \\
\textbf{Global North (N)} & 74 & 16 & 69 & 10 & 0 & 5 & 1 & 2 & 1 & 2 & 90 & 0 & 92 \\
\hline
\end{tabular}
\label{table:demo_intersection}
\end{table}

\section{Results}

\subsection{Collinearity \& Interaction}
\label{app:correlation}
\begin{figure}[h]
    \centering
    \includegraphics[width=0.5\linewidth]{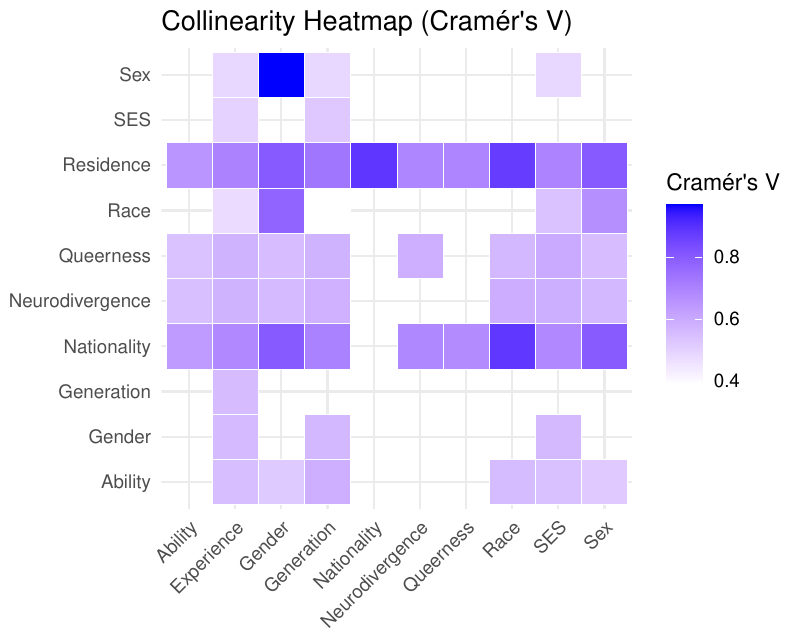}
    \caption{Collinearity heatmap depicting the relationships between demographic variables measured by Cramér's V. Higher values indicate stronger associations between variables.}
    \label{fig:collinearity_heatmap}
\end{figure}
\begin{table}[h]
\caption{Summary of Kruskal-Wallis Interaction Tests}
\begin{tabular}{llrr}
\textbf{Question} & \textbf{Comparison} & \multicolumn{1}{l}{\textbf{Statistic}} & \multicolumn{1}{l}{\textbf{p\_value}} \\ \hline
Q1.1 & Experience - Residence & 20.246918 & 0.0164 \\
Q1.1 & Gender - Nationality & 14.522076 & 0.0023 \\
Q1.1 & Gender - Residence & 12.526991 & 0.0058 \\
Q1.1 & Neurodivergence - Nationality & 17.137 & 0.0042 \\
Q1.1 & Neurodivergence - Residence & 14.129383 & 0.0148 \\ \hline
Q2.1 & Experience - Residence & 21.867787 & 0.0093 \\
Q2.1 & Gender - Nationality & 19.861633 & 0.0002 \\
Q2.1 & Gender - Residence & 16.328727 & 0.0010 \\
Q2.1 & Neurodivergence - Nationality & 19.000754 & 0.0019 \\
Q2.1 & Neurodivergence - Residence & 13.461655 & 0.0194 \\
Q2.2 & Gender - Residence & 9.639053 & 0.0219 \\
Q2.4 & Gender - Nationality & 15.233917 & 0.0016 \\
Q2.4 & Gender - Residence & 16.302808 & 0.0010 \\
Q2.4 & Neurodivergence - Residence & 17.525503 & 0.0036 \\
Q2.6 & Neurodivergence - Nationality & 12.759697 & 0.0257 \\
Q2.7 & Neurodivergence - Nationality & 12.005037 & 0.0347 \\\hline
Q3.1 & Gender - Nationality & 18.053828 & 0.0004 \\
Q3.1 & Gender - Residence & 14.582185 & 0.0022 \\
Q3.1 & Neurodivergence - Nationality & 19.723627 & 0.0014 \\
Q3.1 & Neurodivergence - Residence & 12.856937 & 0.0248 \\
Q3.2 & Gender - Nationality & 10.309245 & 0.0161 \\
Q3.2 & Gender - Residence & 8.270721 & 0.0407 \\
Q3.3 & Gender - Nationality & 15.622442 & 0.0014 \\
Q3.3 & Gender - Residence & 20.203246 & 0.0002 \\
Q3.3 & Neurodivergence - Nationality & 15.437002 & 0.0086 \\
Q3.3 & Neurodivergence - Residence & 20.708036 & 0.0009 \\
Q3.3a & Gender - Residence & 18.53853 & 0.0003 \\
Q3.3a & Neurodivergence - Nationality & 18.831604 & 0.0021 \\
Q3.3a & Neurodivergence - Residence & 21.600736 & 0.0006 \\\hline
\end{tabular}
\label{table:kruskal_interaction}
\end{table}

\begin{table}[]
\caption{Summary of Dunn's Interaction Tests}
\begin{tabular}{llrrr}
\textbf{Question} & \textbf{Comparison} & \multicolumn{1}{l}{\textbf{Z}} & \multicolumn{1}{l}{\textbf{P.unadj}} & \multicolumn{1}{l}{\textbf{P.adj}} \\ \hline
Q1.1 & Man.Global North (N) - Woman.Global South (N) & 3.221152 & 0.0013 & 0.0077 \\
Q1.1 & Man.Global North (R) - Woman.Global South (R) & 3.419984 & 0.0006 & 0.0038 \\
Q1.1 & 11-20 years.Global North (R) - 2-5 years.Global South (R) & 3.84242 & 0.0001 & 0.0055 \\
Q1.1 & Man.Neurotypical - Woman.Neurotypical & 3.306643 & 0.0009 & 0.0142 \\
Q1.1 & Neurotypical.Global North (N) - Neurotypical.Global South (N) & 3.991839 & 0.0001 & 0.0004 \\
Q1.1 & Neurotypical.Global North (R) - Neurotypical.Global South (R) & 3.604055 & 0.0003 & 0.0019 \\
Q1.1 & Straight.Global North (R) - Straight.Global South (R) & 2.668028 & 0.0076 & 0.0458 \\ \hline
Q2.1 & Man.Global North (N) - Woman.Global South (N) & 4.378787 & 0.0000 & 0.0001 \\
Q2.1 & Man.Global North (R) - Woman.Global South (R) & 3.986303 & 0.0001 & 0.0004 \\
Q2.1 & Man.Neurotypical - Woman.Neurotypical & 3.075544 & 0.0021 & 0.0315 \\
Q2.1 & Neurotypical.Global North (N) - Neurotypical.Global South (N) & 3.440893 & 0.0006 & 0.0035 \\
Q2.1 & Neurotypical.Global North (R) - Neurotypical.Global South (R) & 3.019651 & 0.0025 & 0.0152 \\
Q2.1 & Queer.Global South (N) - Straight.Global North (N) & -2.825365 & 0.0047 & 0.0283 \\
Q2.1 & Straight.Global North (N) - Straight.Global South (N) & 3.885102 & 0.0001 & 0.0006 \\
Q2.1 & Straight.Global North (R) - Straight.Global South (R) & 3.160359 & 0.0016 & 0.0095 \\
Q2.2 & Woman.Global North (R) - Woman.Global South (R) & 2.686243 & 0.0072 & 0.0434 \\
Q2.4 & Man.Global North (N) - Woman.Global South (N) & 3.600112 & 0.0003 & 0.0019 \\
Q2.4 & Man.Global North (R) - Woman.Global South (R) & 4.010696 & 0.0001 & 0.0004 \\
Q2.4 & Man.Neurotypical - Woman.Neurotypical & 3.175381 & 0.0015 & 0.0224 \\ \hline
Q3.1 & Man.Global North (N) - Woman.Global South (N) & 4.015621 & 0.0001 & 0.0004 \\
Q3.1 & Man.Global North (R) - Woman.Global South (R) & 3.767677 & 0.0002 & 0.0010 \\
Q3.1 & Man.Neurotypical - Woman.Neurotypical & 3.764253 & 0.0002 & 0.0025 \\
Q3.1 & Neurotypical.Global North (N) - Neurotypical.Global South (N) & 3.866724 & 0.0001 & 0.0007 \\
Q3.1 & Neurotypical.Global North (R) - Neurotypical.Global South (R) & 2.990674 & 0.0028 & 0.0167 \\
Q3.2 & Man.Global North (N) - Woman.Global South (N) & 2.955706 & 0.0031 & 0.0187 \\
Q3.2 & Man.Global North (R) - Woman.Global South (R) & 2.826204 & 0.0047 & 0.0283 \\
Q3.3 & Man.Global North (N) - Woman.Global South (N) & 3.767702 & 0.0002 & 0.0010 \\
Q3.3 & Man.Global North (R) - Woman.Global South (R) & 4.25852 & 0.0000 & 0.0001 \\
Q3.3 & Neurodivergent.Global South (N) - Neurotypical.Global North (N) & -2.699392 & 0.0069 & 0.0417 \\
Q3.3 & Neurodivergent.Global South (R) - Neurotypical.Global North (R) & -2.790012 & 0.0053 & 0.0316 \\
Q3.3 & Neurotypical.Global North (R) - Neurotypical.Global South (R) & 2.738803 & 0.0062 & 0.0370 \\
Q3.3a & Man.Global North (R) - Woman.Global South (R) & 3.185922 & 0.0014 & 0.0087 \\
Q3.3a & Woman.Global North (R) - Woman.Global South (R) & 3.188168 & 0.0014 & 0.0086 \\
Q3.3a & Man.Global South (R) - Woman.Global North (R) & -2.785666 & 0.0053 & 0.0321 \\
Q3.3a & Neurotypical.Global North (N) - Neurotypical.Global South (N) & 2.853872 & 0.0043 & 0.0259 \\
Q3.3a & Neurotypical.Global North (R) - Neurotypical.Global South (R) & 3.103509 & 0.0019 & 0.0115 \\ \hline
\end{tabular}
\label{table:interaction_dunn}
\end{table}

\end{document}